\documentclass{tPHM2e}

\newcommand{\refeq}[1]{Eq.\ (\ref{#1})}  
\newcommand{\refsec}[1]{Sec.\ \ref{#1}}  
\newcommand{\reffig}[1]{Fig. \ref{#1}}   

\newcommand{\ket}[1]{\hbox{$\mid \! {#1} \rangle$}}

\newcommand{\expect}[1]{\hbox{$\langle #1 \rangle$}}
\newcommand{\bigbraket}[3]{\hbox{$\langle \, #1 \, | \, #2 \, | \, #3 \, \rangle$}}

\newcommand{\Tr}{\mathop{\mathrm{Tr}}}

\begin{document}

\title{Schur forms of Matrix Product Operators in the infinite limit}

\author{L. Michel and I. P. McCulloch$^{\ast}$\thanks{$^\ast$Corresponding author. Email: ianmcc@physics.uq.edu.au}\\
  {\em School of Physical Sciences, the University of Queensland, QLD 4072, Australia}}

\maketitle

\begin{abstract}
  Matrix Product State (MPS) wavefunctions have many applications in quantum information and condensed matter
  physics. One application is to represent states in the thermodynamic limit directly, using a small set of
  position independent matrices. For this infinite MPS ansatz to be useful it is necessary to be able to
  calculate expectation values, and we show here that a large class of observables, including operators
  transforming under lattice translations as eigenstates of arbitrary momentum $k$, can be represented in the
  Schur form of a lower or upper triangular matrix and we present an algorithm for evaluating such expectation
  values in the asymptotic limit. The sum or the product of two such Schur operators is also a Schur operator,
  and is easily constructed to give a simple method of constructing arbitrary polynomial combinations of
  operators.  Some simple examples are the variance $\langle (H-E)^2\rangle$ of an infinite MPS, which gives a
  simple method of evaluating the accuracy of a numerical approximation to a eigenstate, or a vertex operator
  $\langle c^\dagger_{k_1} c^\dagger_{k_2} c_{k_4} c_{k_3}\rangle$.  This approach is a step towards improved
  algorithms for the calculation of dynamical properties and excited states.
\end{abstract}


\section{Introduction}

The Matrix Product State (MPS) ansatz \cite{MPS,AKLT} forms the basis of many numerical algorithms, notably
the Density Matrix Renormalization Group (DMRG) \cite{White,UliReview},
and Time Evolving Block Decimation (TEBD) \cite{TEBD}. These algorithms can be applied directly in the
thermodynamic limit of a translationally invariant system (invariant under translations of some fixed
number of lattice sites) \cite{iTEBD,PWFRG,iDMRG}, which gives advantages over traditional finite-size
scaling calculations. In the infinite size variant of DMRG, the converged fixed point produces a
translationally invariant MPS, as studied for example by \"Ostlund and Rommer \cite{OstlundRommer},
which gives a compact representation of an infinite size wavefunction, from which the behavior of
possible correlation functions can be obtained from the spectrum of the transfer operator.
Imaginary time evolution via the iTEBD algorithm \cite{iTEBD} produces the same fixed point, and indeed
the iTEBD and iDMRG algorithms are very similar \cite{iDMRG}, the main difference being the algorithm
for the local update of the tensors in the MPS; whereas TEBD uses a local evolution of a 
single bond, the iDMRG algorithm uses a very efficient local optimization of the total energy. A drawback
of the iTEBD approach is the 
use of the Trotter-Suzuki decomposition \cite{TrotterSuzuki}, which gives a constraint on the size of the 
unit cell which must be a multiple of 2, and interactions beyond nearest neighbor are more difficult,
with typical implementations using relatively inefficient swap gates \cite{TEBDSwapGates}.
Even for nearest-neighbor interactions only, the computation time is linear in the size of the unit cell.
On the other hand, iDMRG allows any size unit cell that is compatible with the periodicity of the
wavefunction, and the performance of the algorithm is essentially independent of the size of the unit
cell, and depends principally on the number of local optimization steps which can be interpreted as
the total size of the lattice (which, for good convergence, will scale with the 
longest correlation length in the system). Nevertheless, (i)TEBD does have some advantages.
As described in \cite{SierraVariational,iDMRG} the 2-site DMRG algorithm introduces a small perturbation
at the location of the bare sites, which is a consequence of a non-zero truncation error.
This causes the variational state to be less than optimal for a given basis size. The effect is small,
nevertheless for approaches based on scaling with respect to the basis size it is an unwanted effect
and the remedy is to approach the converged fixed point in a gradual way, which corresponds to 
imaginary time evolution with a small time-step (this can of course be performed in
DMRG, without a Trotter-Suzuki decomposition, by replacing the eigensolver with a multiplication by $1-\epsilon H$).
For real-time evolution, the time-dependent DMRG algorithm  \cite{UliTime,WhiteTime} 
is indistinguishable from (i)TEBD, save for an unimportant change in notation.

For a homogeneous system, an infinite MPS offers many advantages over a finite-size MPS. The absence of
boundaries avoids many of the problems with Friedel oscillations that complicate the calculation of
correlation functions in finite-size DMRG \cite{NoackFriedel}. On the other hand, conventional finite-size
scaling with respect to the lattice size is not possible, but instead this can be replaced by a scaling
with respect to the largest correlation length in the system $\xi$, which for a critical state scales with
the number of states kept $m$ in the MPS auxiliary basis with a power law \cite{NishinoScaling,LucaScaling},
\begin{equation}
\xi = m^\kappa \; .
\end{equation}
The exponent $\kappa$ is a function only of the central charge of the conformal field theory describing the
critical point \cite{Pollmann}.

The evaluation of local or finite-range expectation values on an infinite MPS is a straightforward calculation. 
On the other hand, in \cite{iDMRG}, the general approach for calculating infinite sums of local terms on an infinite MPS
was presented, for example to compute the energy per site and the fixed point matrices of the Hamiltonian
operator. In this paper, we extend these results to present, in detail, an algorithm for constructing
the expectation values and auxiliary matrices of an arbitrary polynomial function of such operators, 
which includes operators at non-zero momentum, fermionic operators, and string operators.
This generalizes the results presented in \cite{OstlundRommer} for two- and three-body operators.
After discussing some background on matrix product states in the infinite size limit, we
describe in detail the algorithm for obtaining expectation values of triangular MPO's in \refsec{sec:Algorithm},
and as a simple example of these techniques we discuss the utility of using the variance $\sigma^2$ of the energy
as a convergence measure in numerical MPS algorithms in \refsec{sec:Variance}. Finally, we summarize the results
and give some concluding remarks.

\section{Infinite size MPS}

A position independent MPS on an infinite lattice is represented by the form
\begin{equation}
\sum_{\{s_i\}} \cdots A^{s_1} A^{s_2} \cdots
\quad \ket{s_1} \otimes \ket{s_2} \otimes \cdots \; .
\label{eq:MPWavefunction}
\end{equation}
The local index $s_i$ represents an element of the $d$-dimensional local Hilbert space at site $i \in \mathbb{Z}$
of the infinite lattice, and the matrices $A^{s}$ have dimension $m \times m$. Note that sometimes in the literature 
$\chi$ or $D$ is used instead of $m$. In general, we need not require that the unit cell of the state is exactly
one lattice site, but we can have in principle any finite periodicity. Although we present all results here for
the case of a 1-site unit cell, all of the results presented in this paper
generalize straightforwardly to the case of a multi-site unit cell, at a computational cost that is linear
in the size of the unit cell. MPS wavefunctions of this form \refeq{eq:MPWavefunction} have been studied extensively
in the literature, eg in \cite{MPS} and \cite{OstlundRommer}. 
For an introduction to the algorithms for computing the elements of
the infinite MPS representation numerically, see Ref.~\cite{iTEBD,iDMRG}.
In \cite{OstlundRommer}, the MPS ansatz was generalized to a simple representation
for an excited state (Bloch state), which is a generalization of the single-mode approximation (SMA),
constructed by inserting an additional matrix at all possible positions in the lattice,
which we write here in the limit of an infinite size lattice,
\begin{equation}
\ket{Q,k} = \sum_{\{s_i\}} \sum_j e^{ikj} A^{s_1} \cdots A^{s_{j-1}} Q A^{s_j}  A^{s_{j+1}} \cdots \quad \ket{s_1} \cdots
\ket{s_{j-1}} \ket{s_j} \ket{s_{j+1}} \cdots \; .
\end{equation}
This can be represented in a more compact form as a triangular MPS, closely related to the $W$-state
as a position independent MPS with matrices $A^{'s}$ of dimension $2m \times 2m$, given by,
\begin{equation}
A^{'s} = \left( \begin{array}{ll} A^s & 0 \\ QA^s & e^{ik} A^s \end{array} \right) \; .
\end{equation}
In this formulation the final wavefunction is accumulated in the bottom-left matrix entry, rather than on the the diagonal. 
This is a consequence of
the explicit breaking of $U(1)$ charge symmetry to generate a particle-like excitation, thus a finite-dimensional
MPS representation that transforms as a $U(1)$ invariant (scalar) MPS is not possible \cite{MPSRepresentations}.
This form is closely related to the application of a momentum $k$ triangular operator to a state,
for example the matrix product operator (MPO) \cite{DMRG-MPS},
\begin{equation}
B^\dagger_k = \left( \begin{array}{ll} I & 0 \\ b^\dagger & e^{ik}I \end{array} \right) \; ,
\end{equation}
which constructs a boson of momentum $k$, where $b^\dagger$ is the boson creation operator and
$I$ is the identity operator for a single site of the lattice.

A key advantage of the MPO formulation \cite{DMRG-MPS} compared with the \textit{ad hoc} methods traditionally employed
in matrix product numerical approaches, is that the MPO easily allows arithmetic operations, making the construction of
complex operators rather simple. In particular, sums and products of MPO's are constructed simply by taking the matrix direct
sum and matrix direct product respectively of the MPO matrices, and often the resulting MPO can be factorized to
reduce the matrix dimension. For example, an MPO representation of the number operator
$N_k = B^\dagger_k B^{}_k$ is,
\begin{equation}
N_k = \left( \begin{array}{llll} 
I & & & \\ 
b^\dagger & e^{ik}I & & \\ 
b & 0 & e^{-ik}I & \\
b^\dagger b & e^{ik} b & e^{-ik} b^\dagger & I
\end{array} \right) \; .
\end{equation}
The MPO formulation also allows for a convenient and efficient representation of longer range interactions,
as a sum of terms that decay exponentially with distance \cite{Crosswhite}. A surprisingly small number of such terms
can be used to approximate a long-range polynomial decaying interaction.
For a finite system, the expectation value of a triangular MPO operator will be a function system length
that may have a complicated short-range behaviour.
In this paper we calculate the asymptotic functional form
of the expectation value in the limit as we approach an infinitely large system, which approaches a polynomial function.
To see how this works, consider a simple example of the Hamiltonian operator, for example for the Ising model in a transverse
field,
\begin{equation}
H = \sum_i \sigma^z_i \sigma^z_{i+1} + \lambda \sum_i \sigma^x_i \; ,
\end{equation}
which has the MPO representation
\begin{equation}
H = \left( \begin{array}{lll} I & & \\ \sigma^z & 0 & \\ \lambda \sigma^x & \sigma^z & I \end{array} \right) \; .
\end{equation}
On an infinite lattice this expectation value diverges, with the physically relevant quantity being the
energy per lattice site. Equivalently, we can consider an expectation value on $n$ sites of the lattice,
which we denote $\expect{H} {}_n$, which in this example is equal to $nE_0$, where $E_0$ is the energy per site.
More generally, the expectation value of an arbitrary triangular (Schur) operator will be some polynomial in $n$.

On a section of the lattice of length $n$, the exact expectation value of the MPO is not well defined unless a boundary condition
is specified. For example, with open boundary conditions and the Ising Hamiltonian, 
the $\sigma^x_i$ term occurs exactly $n$ times, whereas the nearest-neighbor term $\sigma^z_i \sigma^z_{i+1}$ occurs only
$n-1$ times. This is a consequence of the boundary condition, and in the asymptotic limit the resulting constant term 
is not relevant and in fact is problematic when constructing higher order operators, where such boundary contributions can
potentially affect all sub-leading terms in the polynomial form. These spurious terms are easy to eliminate however,
by constructing the correct fixed point equations for the expectation value. This procedure is described below.

\section{Fixed point equations of triangular MPO's}
\label{sec:Algorithm}

For each index of the $M \times M$ dimensional MPO, we can associate a matrix $E$ (in DMRG notation, this
is called the \emph{block operator}).  To find the expectation value of these operators, we can use a
recursive formula based on the notion that the operators $E_i$ are a function only of the previously
calculated $E_j$, for $j > i$, because of the specific triangular form of the MPO. This gives a recursive algorithm
for the expectation value, whereby we calculate $E_M$, $E_{M-1}$, $\ldots$, $E_1$ in turn.
Solving each $E_i$ matrix will require $O(d m^3)$ operations, giving a total computational complexity of $O(m^3 d M^2)$.

We define the \emph{transfer operator}, which acts on the $m \times m$ $E$ matrices, 
\begin{equation}
  T(E) = \sum_s A^{s\dagger} E A^s \; .
\end{equation}
This operator has a spectral radius of 1, and the eigenspectrum determines the scaling form of all
possible correlation functions \cite{OstlundRommer}.
If the wavefunction is both parity and time inversion invariant, then we can choose a normalization such that
the $A^s$ matrices are symmetric, 
in which case the transfer operator is also symmetric. If the wavefunction is invariant under only
the combination of parity and time ($PT$), then the $A^s$
can be chosen to be Hermitian, in which case the transfer operator is Hermitian.
For a wavefunction that is only $CP$ or $CPT$ symmetric, the $A^s$ matrices can be chosen to be 
symmetric or Hermitian in combination with a
local basis transformation (corresponding to charge inversion), in which case the transfer operator
is not Hermitian, but is a normal operator.
We assume that there is one eigenvalue of $T$ equal to 1. This follows from the normalization of the
wavefunction \cite{OstlundRommer,DMRG-MPS}, and corresponds to the left/right eigenpair of the identity operator
and reduced density matrix. In general there may exist more than one eigenvalue equal to 1, which signals
long range correlations in the state. For the exposition of the procedure for calculating expectation values,
we assume that there is only a single eigenvalue 1, but the generalization to long range correlated states
is straightforward and is discussed in \refsec{sec:AdditionalEigenvalues}.

Generalizing this transfer operator, we can let $X$ be an operator acting on the local Hilbert space of the
MPS, and define
\begin{equation}
  T_X(E) = \sum_{s's} \bigbraket{s'}{X}{s} A^{s'\dagger} E A^s \; .
\end{equation}
Now given an $M \times M$ dimensional MPO $W$, we can express the action of adding one site to the expectation value
in terms of the polynomial form for the $M$ different matrices $E_i$, $1 \leq i \leq M$, as
\begin{equation}
  E_i(n+1) = T_{W_{ii}}(E_i(n)) + \sum_{j>i} T_{W_{ji}}(E_j(n)) \; .
  \label{eq:MPORecurrence}
\end{equation}
The reason why we have split this into two terms, with the diagonal part $W_{ii}$ and the off-diagonal part
$W_{ij}$, is that the $E_j$ matrices for $j>i$ are assumed to be already calculated, so the off-diagonal part
is some matrix function of $n$. Let $C_i(n) = \sum_{j>i} T_{W_{ji}}(E_j(n))$ be the fixed right hand side, and 
let $X = W_{ii}$ be the diagonal element
of the MPO, that acts on the local Hilbert space. Then \refeq{eq:MPORecurrence} reduces to
\begin{equation}
  E_i(n+1) = T_{X}(E_i(n)) + C_i(n) \; .
  \label{eq:MPORecurrenceNicer}
\end{equation}
The operator representing the observable is $E_1(n)$, which has an expectation value of
$\Tr \rho E_1(n)$, where $\rho$ is the reduced density matrix (the right eigenvector of the transfer operator $T$
with eigenvalue 1).
To solve these equations in the large $n$ asymptotic limit we consider several cases, 
firstly zero momentum and later we generalize to
non-zero momentum and string operators.

\subsection{Zero-momentum}

An MPO containing only zero momentum components is characterized by the diagonal components $W_{ii}$ being
proportional to the identity operator, $W_{ii} = xI$, with the prefactor $x$ satisfying either $x=1$ or $|x| < 1$. 
In this case, the $C$ and $E$ matrices are polynomial functions of $n$,
with matrix-valued coefficients. Therefore, let
\begin{equation}
  \begin{array}{rcl}
    C(n) &=& \sum_{m=0}^p C_m n^m \\
    E(n) &=& \sum_{m=0}^{p+1} E_m n^m
  \end{array}
\end{equation}
where $C_m$ and $E_m$ are matrix-valued coefficients of the $p$ and $p+1$ degree polynomials $C(n)$ and $E(n)$ respectively.
For clarity of notation we have suppressed the subscript $i$ from the $C(n)$ and $E(n)$ matrices, with the understanding
that the same form of fixed point equations will be solved for each column of the MPO.
Now let diagonal operator $W_{ii} = xI$, where $x$ is a $c$-number. We can further divide into sub-cases:
If $x=0$, then \refeq{eq:MPORecurrenceNicer} reduces to simply $E(n+1) = C(n)$, or equating
coefficients,
\begin{equation}
E_m = C_m - \sum_{j=m+1}{p} \binom{j}{m} E_j \; ,
\label{eq:MPORecurrenceTrivial}
\end{equation}
which again is obtained recursively starting from $E_p$, $E_{p-1}$, $\ldots$, $E_1$.

The second case is $|x| < 1$. In this case, \refeq{eq:MPORecurrenceNicer} reduces to
\begin{equation}
  E_i(n+1) = x T(E_i(n)) + C(n) \; .
\end{equation}
Equating coefficients of the polynomial expansion, we get
\begin{equation}
  (1-xT)(E_m) = C_m - \sum_{j=m+1}^{p+1} \binom{j}{m} E_j \; .
\label{eq:MPORecurrenceLinear}
\end{equation}
For each index $m = p, p-1, \ldots, 0$, the right hand side is a fixed matrix so the $E_m$ is obtained as the solution
of a set of linear equations. The solution to these equations corresponds to taking the limit $n \rightarrow \infty$, such
that the geometric series defined by \refeq{eq:MPORecurrenceNicer} converges to a fixed point.
Since we have $|x| < 1$, the operator $1-xT$ is non-singular and the solution is unique.
If the transfer operator $T$ is Hermitian, then this set of linear equations can be solved using a simple conjugate gradient
method. For more general cases, conjugate gradient is not suitable but the
GMRES algorithm gives good convergence at the cost of higher memory requirements, although this is typically not
a significant limitation in this context.

The final case for zero-momentum operators is when the prefactor $x=1$. Because the transfer operator $T$ has an
eigenvalue 1, this means that the left hand side of \refeq{eq:MPORecurrenceLinear} is singular
and it is necessary to decompose the matrices $E(n)$ and $C(n)$ into components parallel and perpendicular to the
identity (ie.~ the eigenvector corresponding to the eigenvalue 1 of $T$). We do this via
\begin{equation}
  \begin{array}{rcl}
    E_m &=& e_m I + \bar{E}_m \\
    C_m &=& c_m I + \bar{C}_m
  \end{array}
\end{equation}
where $\bar{E}_m$ denotes the component of $E_m$ perpendicular to the identity operator and $e_m$ is the coefficient
of the component of $E_m$ in the direction of the identity operator. Now, the components $\bar{E}_m$ are well defined
by \refeq{eq:MPORecurrenceLinear}, since these matrices are orthogonal to the singular component of $1-T$.
In a numerical implementation it is necessary to take special care however to ensure that the solution to the linear
equation is stable, by removing any spurious components in the direction of the identity during the course of
the solver algorithm.

The components in the direction of the identity satisfy
\begin{equation}
  e_{m+1} = \frac{1}{m+1} \left[ c_m - \sum_{k=m+2}^{p+1} \binom{k}{m} e_k \right] \; ,
  \label{eq:E1Ident}
\end{equation}
which again can be solved straightforwardly starting from $m=p$ down to $m=0$. 
Hence if
$c_p \neq 0$, then the degree of the polynomial will be increased by 1, as we will end up with a non-zero
component $e_{p+1}$.  Note also that $e_0$ is not defined in this procedure, which is what we expect since the
constant offset of the expectation value is a boundary term. We are free to choose $e_0 = 0$,
which is the choice that removes spurious sub-leading boundary contributions.

\subsection{Finite momentum}

An MPO involving operators of finite momenta introduces oscillating components into the expectation value, that
cannot be represented by a polynomial. However, as long as there is only a \emph{finite} number of momenta,
which will always be true for a finite-dimensional MPO, we can write each coefficient matrix $E(n)$ and $C(n)$ 
uniquely as
a sum over momenta, as
\begin{equation}
  \begin{array}{rcl}
    C(n) &=& \sum_k e^{ikn} C^{(k)}(n) \\
    E(n) &=& \sum_k e^{ikn} E^{(k)}(n)
  \end{array}
\end{equation}
where the $k$ summation is over all distinct momenta with non-zero contributions.
We can then expand the $C^{(k)}(n)$ and $E^{(k)}(n)$ as polynomials in $n$ as before, and equate coefficients
of $e^{ikn} n^m$ for each $k,m$.
We now consider each of the possible cases for the diagonal operator $X = W_{ii}$, generalizing the results from
the previous section.

Firstly, if $X=0$, then \refeq{eq:MPORecurrenceTrivial} acquires a phase factor, giving
\begin{equation}
  E^{(k)}_m = e^{-ik} C^{(k)}_m - \sum_{j=m+1}{p} \binom{j}{m} E^{(k)}_j \; ,
  \label{eq:MPORecurrenceTrivialK}
\end{equation}
and similarly, when $X = xI$ with $|x| < 1$, we get a phase factor difference from \refeq{eq:MPORecurrenceLinearK}, giving
\begin{equation}
  (1-e^{-ik} xT)(E_m) = e^{-ik} C_m - \sum_{j=m+1}^{p+1} \binom{j}{m} E_j \; .
  \label{eq:MPORecurrenceLinearK}
\end{equation}
Now the third case, for $X = xI$ with $|x|=1$, let $x = e^{ik}$. We have two distinct possibilities.
If there is a non-zero component $C^{(k)}_m$ with the same momentum, then we get a diverging component and
we must again treat the components parallel and perpendicular to the identity (eigenvector of the transfer operator $T$
with eigenvalue 1) separately. For the perpendicular components $\bar{E}^{(k)}_m$, we get
\begin{equation}
  (1-T)(\bar{E}^{(k)}_m) = e^{-ik} \bar{C}^{(k)}_m - \sum_{j=m+1}^{p+1} \binom{j}{m} \bar{E}^{(k)}_j \; ,
\end{equation}
while the component in the direction of the identity, $e^{(k)}_m$, is
\begin{equation}
  e^{(k)}_{m+1} = \frac{1}{m+1} \left[ e^{-ik} c^{(k)}_m - \sum_{j=m+2}^{p+1} \binom{j}{m} e^{(k)}_j \right] \; .
    \label{eq:E1IdentK}
  \end{equation}
For components $C^{(k')}_m$ with a different momentum $k' \neq k$, the components perpendicular to the identity converge,
\begin{equation}
  (1-e^{i(k'-k)}T)(\bar{E}^{(k')}_m) = e^{-ik'} \bar{C}^{(k')}_m - \sum_{j=m+1}^{p+1} \binom{j}{m} \bar{E}^{(k')}_j \; ,
\label{eq:IdentKperp}
\end{equation}
while the component in the direction of the identity acquires a new component at the momentum $k$,
\begin{eqnarray}
  e^{(k')}_m &=& \displaystyle \frac{c^{(k')}_m}{e^{ik'} - e^{ik}} \\
  e^{(k)}_m &=& \displaystyle \frac{-c^{(k')}_m}{e^{ik'} - e^{ik}}
\label{eq:IdentKp}
\end{eqnarray}
This applies in an additive sense, so that if there are many different components $c^{(k')}_m$ with different momenta $k'$,
the coefficient $e^{(k)}_m$ is the sum of all $ \frac{-c^{(k')}_m}{e^{ik'} - e^{ik}}$.

For the finite momentum case, the final matrix $E_1(n)$ will contain in general many oscillating components with
different momenta $k$. The actual expectation value is obtained when every term contributes an integer number
of wavelengths $\lambda = 2\pi/k$. Therefore, we can restrict $n$ to be the lowest common multiple of all wavelengths
$\lambda$, in which case all of the oscillating components $e^{ikn}$ become equal to 1 and we recover a simple polynomial.
In fact, this procedure also works if the $\lambda$ are irrational, in the sense that for a large enough $n$ we can
get all components arbitrarily close to containing an integer number of wavelengths.

\subsection{String operators}
\label{sec:AdditionalEigenvalues}

Now that we have the machinery to handle non-zero momenta, the generalization to operators on the diagonal $W_{ii} = X$
which are \emph{not} proportional to the identity operator is straightforward, and in this section we sketch the solution.
Applications where non-trivial operators occur on the diagonal of a triangular MPO include string order parameters,
where the diagonal component is some unitary operator, and fermionic operators, where the diagonal component will be the local
number parity operator $(-1)^N$, corresponding to a Jordan-Wigner string.
We can assume that it is normalization of the operator is such that the
spectral radius of the transfer operator $T_X$ is at most 1 (otherwise the expectation value may diverge exponentially).
We can easily relax the requirement that there is at most one eigenvalue of $T_X$ equal to 1, which
covers the case where the wavefunction contains long range correlations.
To solve the fixed point equations for such an operator, we must identify the eigenvector subspaces of $T_X$ with eigenvalue 1
and eigenvalues of the form $e^{ik}$, with norm 1.
The components of the $E_i(n)$ operators orthogonal to these subspaces converge to a fixed point in the same way as
we have treated previously, according to
\refeq{eq:MPORecurrenceTrivialK}, \refeq{eq:MPORecurrenceLinearK} and \refeq{eq:IdentKperp}. For each component in
the direction of an eigenvector of $T_X$ with eigenvalue of norm 1, we treat the coefficient in the same way as for
the components in the direction of the identity operator, \refeq{eq:E1IdentK} and \refeq{eq:IdentKp} which must be done
separately for each such component. 

\section{Variance as a test of convergence}
\label{sec:Variance}

A simple application of this method is to calculate the expectation value of the square of the Hamiltonian operator.
This gives a polynomial of degree 2, and for the typical case where there are no long range correlations
in the wavefunction the coefficient of the $n^2$ term is simply the square of the groundstate energy per site.
Of more interest is the linear term, which gives the variance per site,
\begin{equation}
\expect{(H-nE_0)^2}_n = \expect{H^2}_n - n^2E_0 = n \sigma^2 \; ,
\end{equation}
where $E_0$ is the energy per site.
The variance gives a simple and reliable measure of how close a given wavefunction is to being an eigenstate of $H$.
In DMRG calculations the truncation error is usually used for this purpose \cite{UliReview}, however the truncation error
is not an intrinsic property of the wavefunction itself but rather is a byproduct of the particular choice of algorithm.
Indeed, for some variants of MPS algorithms, such as single-site DMRG and TEBD in the limit of small time-step, the
truncation error is identically zero and is therefore not useful. The variance however is an observable that can be used on
any MPS wavefunction, irrespective of how the state was originally obtained. To test the utility of this approach,
we have calculated the variance per site in the infinite size limit of the isotropic spin 1/2 Heisenberg model,
\begin{equation}
H = \sum_{<ij>} \left( -S^x_i S^x_j - S^y_i S^y_j + S^z_i S^z_j \right) \; ,
\end{equation}
using iDMRG with a 1-site unit cell. As the number of states is varied, the variance and energy change according
to \reffig{fig:VariancePlot}. Similarly to the well-known case of the truncation error in DMRG \cite{UliReview},
the variational energy is a linear function of the variance $\sigma^2$. For this example
a linear fit using the 10 most accurate data points gives a groundstate energy of $-0.443147181$, 
correct to 9 significant figures compared with
the exact result \cite{ExactXXX} of $1/4 - \ln 2 \simeq -0.4431471805599\ldots$. By comparison, the best variational result in
the calculated data, for 200 states kept, is $-0.44314711$, correct to only 7 significant figures. The accuracy
of the fit is well captured by the standard error of the fit, of $\sigma_E = 2 \times 10^{-9}$, whereas
linear fits calculated via the truncation error give an error that is often somewhat too small. This can
be explained as a systematic error in the DMRG algorithm, by the convergence as the number of basis states is changed
where the truncation error typically changes quite smoothly. The
truncation error measures convergence of the eigenvalues of the reduced density matrix, which can be misleading
because the eigenvalues converge much faster than the eigenvectors. This is the critical difference to the
variance, which instead measures the convergence of the eigenvectors themselves and is therefore a more robust measure of
convergence.

\begin{figure}
\begin{center}
\caption{Energy versus variance in the spin 1/2 Heisenberg model, with a number of states kept varying from $m=40$ to
$m=200$. Inset shows a closeup of the points used to obtain the linear fit.}
\label{fig:VariancePlot}
\end{center}
\end{figure}

\section{Summary and conclusions}
\label{sec:Conclusions}

In this paper, we have developed an algorithm for obtaining the expectation value and corresponding block operators
of arbitrary triangular (Schur) matrix product operators, which are obtained as a polynomial function of the lattice size
$n$, in the large $n$ limit. Operators of this type are useful for many purposes, and we have described a simple example
where the square of the Hamiltonian can be used as an effective convergence test in numerical calculations, superseding
the well-known truncation error from DMRG. In \cite{DMRG-MPS}, one of us showed that for finite size MPS,
it is practical to calculate power series and perturbative expansions to a dozen or so orders, and the algorithm
we have presented here is similarly useful for obtaining the first few terms of an expansion, for example to
obtain the pole expansion of the Green's function $G(w,k)$.
Other applications include obtaining expectation values of operators at
finite momentum directly in the thermodynamic limit, which share a similar structure to the ansatz for an excited state
studied previously by \"Ostlund and Rommer \cite{OstlundRommer}. We expect that this algorithm will be an important component
for constructing improved variational algorithms based around similar excited state ans\"atze.

\section*{Acknowledgements}
Thanks to Guifre Vidal for advice and stimulating conversations. IPM thanks
Halina Rubenzstein-Dunlop, Michael Drinkwater,
Gerard Mulburn, Ross McKenzie, Luca Tagliacozzo and Mark Roberts for support and advice during the preparation of this work.

\end{document}